\documentclass[pra,twocolumn,floats,10pt,floatfix,nobibnotes]{revtex4-1}

\usepackage{amsmath}
\usepackage{amsfonts}
\usepackage{amssymb}
\usepackage{graphicx}
\usepackage{color}
\usepackage[colorlinks=true,citecolor=blue,linkcolor=blue,urlcolor=black]{hyperref}
\usepackage{times}
\usepackage{amstext}
\usepackage{latexsym}
\usepackage{graphicx}% Include figure files
\usepackage{epstopdf}
\usepackage{lipsum}
\usepackage{bbm}
\usepackage[final]{pdfpages} 
\usepackage{etoolbox}
\makeatletter
\patchcmd{\@outputpage@head}{\@ifx{\LS@rot\@undefined}{}{\LS@rot}}{}{}{}
\makeatother

\begin{document}

%\includepdf[pages={1-5}]{SMDDITV5}

%\preprint{APS/123-QED}

\title{Multiple Transparency Windows and Fano interference Induced by Dipole-Dipole Couplings}

\author{E. C. Diniz}
%\email{ediniz98@gmail.com}
\author{H. S. Borges}
\email{halyneborges@df.ufscar.br}
\author{C. J. Villas-Boas}
%\email{villasboas@ufscar.br}
\affiliation{Departamento de F\'{i}sica, Universidade Federal de S\~{a}o Carlos, P.O. Box 676, 13565-905, S\~{a}o Carlos, S\~{a}o Paulo, Brazil}

\begin{abstract}
We investigate the optical properties of a two-level system (TLS) coupled to a linear series of $N$ other TLS's with dipole-dipole coupling between the first neighbours. The first TLS is probed by weak field and we assume that it has a decay rate much stronger than the decay rates of the other TLS's. For N=1 and in the limit of a probe field much weaker than the dipole-dipole coupling, the optical response of the first TLS, i.e., its absorption and dispersion, are equivalent to those of a three-level atomic system in the configuration which allow one to observe electromagnetically induced transparency (EIT) phenomenon. Thus, here we are investigating a new kind of induced transparency where the dipole-dipole coupling plays the same role of the control field in EIT in three-level atoms. We describe this physical phenomenon, here named as Dipole-Dipole Induced Transparency (DDIT), and investigate how it scales with the number of coupled TLS's. In particular we have shown that the number of TLS's coupled to the main one is exactly equals to the number of transparency windows. The ideas presented here are very general and can be implemented in different physical systems such as array of superconducting qubits, array of quantum dots, spin chains, optical lattices, etc.
\end{abstract}

\maketitle

The understanding of the light-matter interaction has been the focus of intense research during the last decades, mainly due to the advances in its manipulation allowed by the introduction of laser fields. Many of these efforts are justified in view of the possibility of using it for the implementation and control of quantum systems on a variety of topics including quantum computing \cite{Kane1998, Ladd2010}, collective atomic phenomena \cite{Gross1982}, trapped ions \cite{Leibfried2003}, cavity and circuit QED \cite{Reiserer2015, Xiang2013}, and other applications involving microscopic scales. Despite the difficulty related to the control and implementation of coupled quantum systems, which are essential for building scalable quantum networks \cite{Duan2010}, significant advances have been achieved using some quantum devices in last years \cite{Reiserer2015, Xiang2013,Duan2010}. In this sense, electromagnetically induced transparency (EIT) \cite{Boller1991, Fleischhauer2005} has been shown to be a phenomenon very useful for manipulating light with light, allowing applications as optical transistor \cite{Parkins2010, Mucke2010}, quantum memories \cite{Choi2008, Specht2011}, to generate  controllable phase shifts on single photon pulses \cite{Borges2016}, ground state cooling of either single atoms \cite{Morigi2000, Roos2000} or ion strings \cite{Lechner2016} among many others. 

The first sistematic experimental study on EIT were carried out with three-level atoms in $\Lambda$ configuration \cite{Boller1991}. In this system, the absorption of a weak probe field, resonant with some atomic transition, can be cancelled out by coupling the excited atomic state to an additional ground state through another laser light, called control field. Due to this coupling, the system presents two different absorption pathways for the probe field, which can interfere whenever the Rabi frenquency of the control field is smaller or of the order of the total decay rate of the excited atomic state \cite{Fleischhauer2005}. Also, some important properties such as the width of the transparency window can be directly adjusted via the Rabi frequency of this control field \cite{Fleischhauer2005}. Since the appearance of the work by Boller \textit{el al.} \cite{Boller1991}, the fundamental idea of the EIT has been extended to other systems. Nowadays we one can observe the interference between different absorption pathways, resulting in adjustable transparency windows, in a large variety of different physical systems as coupled classical \cite{Alzar2002} or quantum harmonic oscillators \cite{Ponte2005}, two-coupled optical cavity modes \cite{Xu2006, Souza2015}, two-level atom coupled to a cavity mode \cite{Rice1996, Waks2006}, quantum dot molecule \cite{Yuan2006,Borges2012, Borges2017}, plasmonic systems \cite{Liu2009, Lu2012}, optomechanical oscilators \cite{Weis2010, Sohail2016}, metamaterials \cite{Papasimakis2008, Tassin2009}, etc.  By applying more control fields, coupling the excited state to additional ground states, more transparency windows can be obtained, thus appearing double-EIT phenomenon, for instance, in a four-level tripod atomic system \cite{Li2007}. Double-transparency windows can also be observed in multiple coupled photonic crystal cavities \cite{Yang2009} or in optomechanical system \cite{Ma2014}. These ideas can also be extended to multiple transparency windows which can be achieved either in $(N+1)$-level atomic system \cite{Paspalakis2002} or in $N$ periodically coupled whispering gallery-mode resonators \cite{Xiao2007}.

In this work we investigate the optical response of a TLS coupled to a series of others TLS's which interact with its first neighbours via dipole-dipole. Our results show that the dipole-dipole coupling plays exactly the same role as the control field in the EIT phenomenon, either in free space \cite{Fleischhauer2005, Souza2015} or in cavity/circuit QED experiments \cite{Mucke2010}. We also investigate the scalability of this system, i.e., how it is possible to control the number of transparency windows. 

\section*{DDIT in Free Space}
\begin{figure}[t]
\centering
\includegraphics[width=0.45\textwidth]{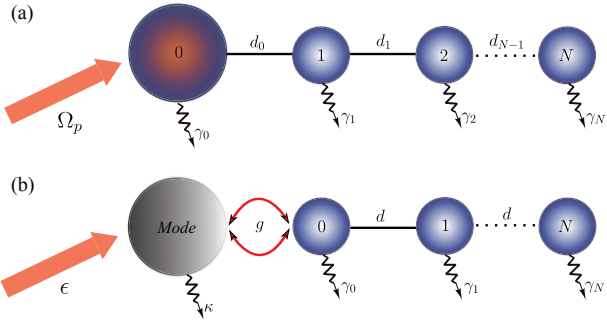}
\caption{Pictorial representation of the system. (a) $1+N$ coupled Two-Level Systems (TLS's), with dipole-dipole coupling $d_i$ between the first neighbours, with the first TLS driven by a coherent (probe) field of Rabi frequency $2\Omega_p$ and oscillating frequency $\omega_p$. (b) A driven cavity mode coupled to $1+N$ TLS's. The coupling between the cavity mode and the first TLS is given by the vacuum Rabi frequency $g$ and the probe field drives the cavity with strength $\epsilon$.}
\label{fig:Scheme}
\end{figure}

Here we analyse the optical response of a two-level system (TLS) driven by a probe field, which in turn is coupled to $N$ others TLS's with dipole-dipole coupling between the first neighbours, as schematically shown in Fig. \ref{fig:Scheme}(a). We assume a weak probe field, oscillation frequency $\omega_p$, driving only the main TLS, such that the Hamiltonian which describes the dynamics of this system is given by ($\hbar=1$)
\begin{equation}
\label{eq1_hamiltonian}
H=\overset{N} {\underset{i=0}{\sum }}\frac{\omega _{0}\sigma _{z}^{i}}{2}%
+\overset{N-1} {\underset{i=0}{\sum }}(d_i \sigma _{-}^{i}\sigma _{+}^{i+1}+ h.c.)+\Omega
_{p}(\sigma _{+}^{0}e^{-i\omega _{p}t}+h.c).
\end{equation}%
Here $\sigma_z^i = (|e \rangle\langle e|^{(i)}-|g \rangle\langle g|^{(i)})$ and $ \sigma_{-}^i = (\sigma_{+}^i)^\dagger = |g \rangle \langle e|^{(i)}$ are the operators (Pauli's matrix) of the $i$-th TLS, being $|e \rangle$ and $|g \rangle$ the excited and ground states, respectively, whose transition frequency is $\omega_0$. $d_i$ is the dipole-dipole coupling, and $2\Omega_p$ is the Rabi frequency of the probe field. The Hamiltonian above can be found/engineered in a large variety of different physical systems such as optical lattices \cite{Duan2003, Garcia-Ripoll2004}, in an array of coupled optical cavities with single trapped atoms inside \cite{Joshi2013}, coupled superconducting qubits \cite{Salathe2015}, in trapped ion domain \cite{Blatt2012}, or in array of quantum-dots \cite{Al-Ahmadi2006, Zajac2016}.  Writing  the Eq.(\ref{eq1_hamiltonian}) in the rotating frame of the driving field we end up with the time-independent Hamiltonian
\begin{equation}
\label{eq1_Hfree}
H=\overset{N} {\underset{i=0}{\sum }}\frac{\Delta _{p}\sigma _{z}^{i}}{2}%
+ \overset{N-1} {\underset{i=0}{\sum }}(d_i \sigma _{-}^{i}\sigma _{+}^{i+1}+ h.c.)+\Omega
_{p}(\sigma _{+}^{0}+h.c),
\end{equation}
where $\Delta _{p}=\omega _{0}-\omega _{p}.$ 

Considering the environment at $T=0K$ and the limit of weak system-reservoir and weak dipole-dipole ($|d_i| \ll\omega_0$) couplings, the dissipation mechanisms of the whole system can be taken into account via the master equation in the Lindblad form \cite{Pettruccione}
\begin{equation}
\label{mater_eq_free}
\frac{d\rho }{dt}=-i\left[ H,\rho \right] +\overset{N} {\underset{i=0}{\sum }} \gamma _{i}(2\sigma _{-}^{i}\rho \sigma _{+}^{i}-\sigma_{+}^{i}\sigma _{-}^{i}\rho -\rho \sigma _{+}^{i}\sigma _{-}^{i}),
\end{equation}%
being $\gamma_i$ the decay rate of the $i$-th TLS. By imposing $d\rho/dt = 0$ we can obtain the steady state and then investigate the optical response of the main TLS such as absorption and dispersion, here defined as $\text{Im}\langle \sigma_{+}^0 \rangle_{ss}$ and $\text{Re}\langle \sigma_{+}^0 \rangle_{ss}$, respectively. The stationary solution for this system can be derived analytically for arbitrary number of TLS's and arbitrary set of parameters. For some special cases, we can obtain simple expressions for $\langle \sigma_{+}^0 \rangle_{ss}$, which allow us to identify important new physical phenomena. For instance, considering only the main TLS coupled to a single other ($N=1$), in the weak probe field limit $|\Omega_p| \ll |d_0|$, and the  main qubit having a stronger decay rate than the other one, i.e., $\gamma_0 \gg \gamma_1$, which allow us to approximate $\langle \sigma^0_{-} \sigma^0_{+} \rangle \approx 1$, we obtain the following stationary solution
\begin{equation}
\label{sigma_p_ss}
\langle \sigma_{+}^0 \rangle_{ss}=Tr(\rho_{ss}\sigma _{+}^{0}) \approx \frac{\left(\Delta_{p}+i\gamma_{1}\right)\Omega_{p}}{|d_0|^{2}-\left(\Delta_{p}
+i\gamma_{0}\right)\left(\Delta_{p}+i\gamma_{1}\right)}.
\end{equation}
From this expression we can straight derive the dispersion and absorption, its real and imaginary parts, respectively, and then we can analyse the optical properties of this system. 

Our system composed by two coupled TLS's must be compared to the system constituted by two quantum dots with dipole-dipole coupling employed to perform an optical switching \cite{Gea-Banacloche2006}. However, different from our system, in \cite{Gea-Banacloche2006} the transparency is not induced by the dipole-dipole interaction and the authors assume two fields (probe and a control) acting simultaneosly on both quantum dots (TLS's) and the same decay rates for them. In this way, they are able to show an efficient optical switching only when the Rabi frequency of the control field is much stronger than the decay rate the quantum dots. In fact, in this regime one has an Autler-Townes splitting instead of a real interference between different excitation pathways, which is the fundamental process behind EIT. 

\begin{figure}[t]
\centering
\includegraphics[width=0.5\textwidth]{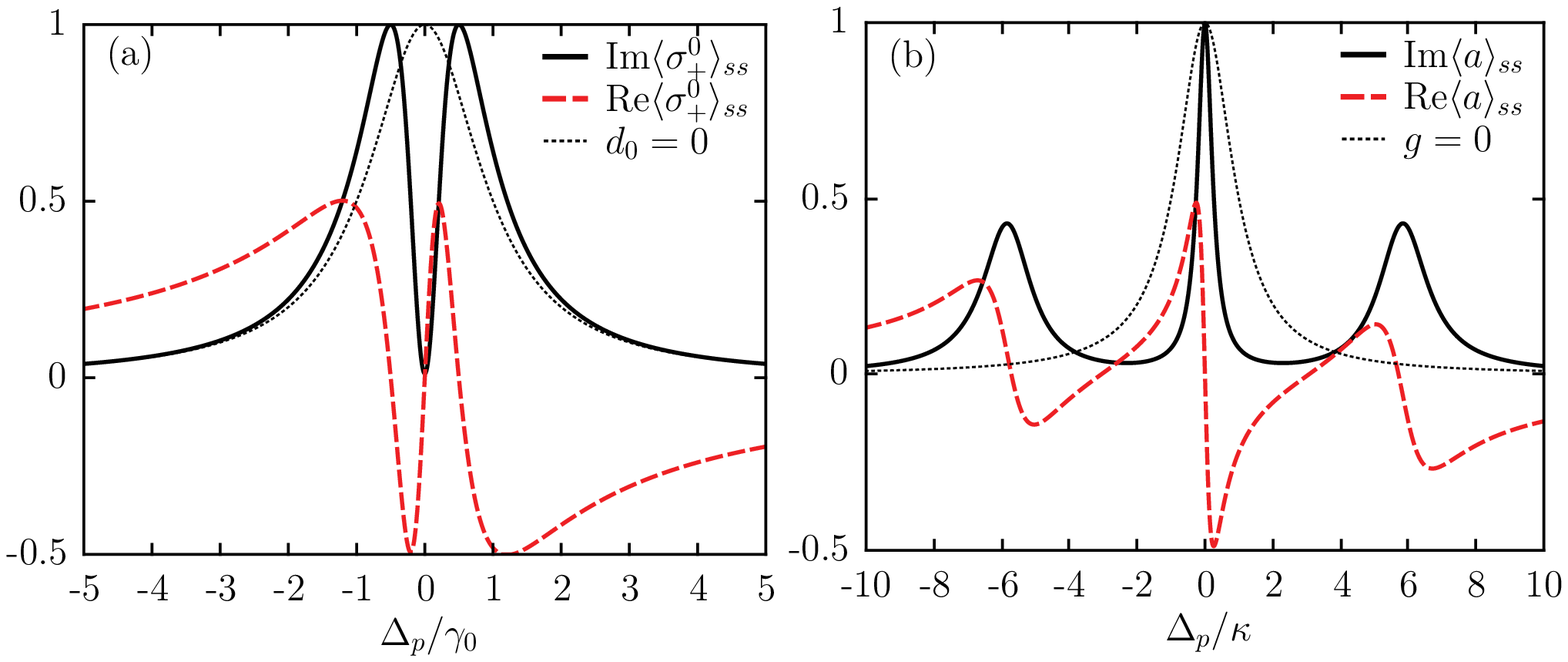}
\caption{(a) Normalized absorption, $\text{Im}\langle \sigma_{+}^0 \rangle_{ss}$, (black solid line) and dispersion, $\text{Re}\langle \sigma_{+}^0 \rangle_{ss}$, (red dashed line) of the first TLS when coupled to a second TLS as a function of the normalized detuning $\Delta_p /\gamma_0$. The parameters used here were $\Omega_p = 0.03 \gamma_0$, $d_0 =0.5 \gamma_0$ and $\gamma_1 = 10^{-3}\gamma_0$. (b) Normalized absorption, $\text{Im}\langle a \rangle_{ss}$, (black solid line) and dispersion, $\text{Re}\langle a \rangle_{ss}$, (red dashed line) of the cavity mode when coupled to two-coupled TLS's as a function of the normalized detuning $\Delta_p /\kappa$. Parameters used: $\gamma_0 = \kappa$, $g = 5\kappa$, $d = 3\kappa$, $|\epsilon| = 0.03\kappa$, and $\gamma_1 = 10^{-3}\kappa$. The black dotted lines represents the absorption when there is no TLS coupled to the main one ($d_0 = 0$) or to the cavity mode ($g=0$).}
\label{fig:ddit_FS_cav}
\end{figure}

Comparing the absorption and dispersion of our system ($\text{Real}\langle \sigma_{+}^0 \rangle_{ss}$ and $\text{Im}\langle \sigma_{+}^0 \rangle_{ss}$) with those of a three-level atomic system in the electromagnetically induced transparency regime \cite{Fleischhauer2005, Souza2015}, we immediately recognize a new kind of induced transparency in which the dipole-dipole coupling $d_0$ plays the same role as the Rabi frequency of the control field. We can also see that the decay rate of the first (second) TLS plays the same role as the total decay rate of the excited state $\Gamma$ (dephasing rate - $\gamma_{ph}$) in three-level systems in EIT regime, which make clear the requirement for different decay rates for the two TLS's employed in our model. Thus, here we have a physical phenomenon which we named as \textit{dipole-dipole induced transparency} (DDIT). In Fig. \ref{fig:ddit_FS_cav}(a) we plot the imaginary (absorption) and real (dispersion) parts of $\langle \sigma_{+}^0 \rangle_{ss}$ for a set of parameters (see figure caption) which allows the observation of DDIT. Keeping $|d_0| \gg |\Omega_p|$ and $|d_0| < \gamma_0$ we can note that the transparency window directly depends on the dipole-dipole coupling $d_0$, as expected. In another related work \cite{Puthumpally-Joseph2014}, the authors claim that it is possible to observe a similar effect, i.e., a dipole induced transparency, in high-density atomic medium which contains two species of atoms (different dipoles). However, as they assume the same decay rate for both dipoles, they can not observe a transparency window extremely narrow as usually allowed in EIT experiments \cite{comment_DIT}.

\section*{DDIT in Cavity/Circuit QED}
Considering a three-level atom, in the EIT configuration, coupled to a cavity mode, one can observe cavity-EIT \cite{Mucke2010}. According to the discussion above, the same effect should be observed by replacing the three-level atom by two coupled TLS's. This is in fact the case, as we will explain below. To this end, firstly let us describe a more general system, i.e., to consider the interaction of a quantum cavity mode with a series of $1+N$ other two-level systems, as schematically represented in Fig. \ref{fig:Scheme}(b). We assume a series of $1+N$ identical TLS's, with dipole-dipole coupling $d$ and individual decay rates $\gamma_i$. The first TLS is then resonantly coupled to the cavity mode, coupling $g$, which is driven by a probe field of strength $\epsilon$ and frequency $\omega_p$. The Hamiltonian of this system in the rotating frame of the probe field reads (for $\hbar = 1$)
\begin{eqnarray} 
\label{Hc}
H_c=\Delta_p a^{\dagger} a + \frac{\Delta_p}{2} \underset{i=0}{\overset{N}{\sum }}\sigma_z^i  + g \left( a\sigma_{+}^0 +h.c. \right) + \nonumber \\
d\underset{i=0}{\overset{N-1}{\sum }}(\sigma_{-}^{i}\sigma _{+}^{i+1}+h.c) + \left(\epsilon a+h.c \right),
\end{eqnarray}
with $\Delta_p = \omega_0 - \omega_p = \omega_{cav}-\omega_p$, being $\omega_{cav}$ the cavity mode frequency. In this case, the dissipation of the cavity mode can be taken into account by adding the term $\kappa(2 a\rho a^{\dagger} - a^{\dagger}a\rho  -\rho a^{\dagger}a)$ into the master equation of the system, being $\kappa$ the decay rate of the cavity field's amplitude. This new master equation can be analytically solved for some particular set of parameters. Considering two TLS's and $|\epsilon/\kappa| \ll 1$, which implies a very small average number of photons inside the cavity mode (whose maximum is given by $|\epsilon/\kappa|^2$) we can derive the steady state solution for the average value of the annihilation operator of the cavity mode (see Supplementary material for the details of its derivation) which reads 
\begin{eqnarray}
\left\langle a\right\rangle_{ss}  &=&\frac{ \epsilon d^{2}-\epsilon \left( \Delta_p-i\gamma _{0}\right) \left( \Delta_p-i\gamma _{1}\right) }{-g^{2}\left( \Delta_p-i\gamma _{1}\right) -d^{2}(\Delta_p-i\kappa )+ \Psi}
\end{eqnarray}
where $\Psi = \left( \Delta_p-i\gamma _{0}\right) \left( \Delta_p-i\gamma _{1}\right) (\Delta_p-i\kappa )$.
This equation must be compared to the equation for the average value of the annihilation operator for a cavity mode coupled to a three-level atom in the EIT configuration (cavity-EIT) \cite{Souza2015}. Again, the dipole-dipole coupling plays exactly the same role as the control field in the cavity-EIT experiments. The width of the EIT resonance (central peak) is proportional to $|d/g|^2$ (see Supplementary material) and then it can be adjusted either via atom-cavity mode or dipole-dipole couplings. In Fig. \ref{fig:ddit_FS_cav}(b) we plot the normalized absorption and dispersion of the cavity mode when coupled to two-coupled TLS's, here defined as $\text{Im}\langle a \rangle_{ss}$ and $\text{Re}\langle a \rangle_{ss}$, respectively.

\section*{Scalability of the system: multiple DDIT}
The results above can be properly extended to multiple transparency windows by adding more TLS's, as schematically shown in Fig. \ref{fig:Scheme}. Thus, from now we will continue to investigate what happens to the optical properties of the system when coupled to a series of other qubits. We assume that the coupling between the first TLS ($i=0$) and the second one is still given by $d_0$ and, for simplicity, the coupling between the other TLS's is $d$, i.e,  $d_1 = d_2=...=d_{N-1} = d$. We also assume the same decay rate for the other TLS's $\gamma_1 = \gamma_2 =...=\gamma_N =\gamma \ll\gamma_0$. 

To understand what happens to the system when we couple more TLS's, it is instructive to analyse the eigenenergies of the bare Hamiltonian (without the probe field).
\begin{figure}[t]
\centering
\includegraphics[width=0.5\textwidth]{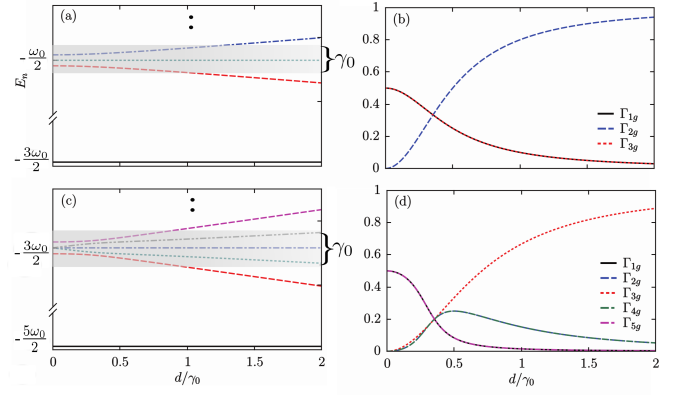}
\caption{Left panels: first eigenenergies (ground state and eigenstates with one excitation) as a function of $d/\gamma_0$. Right panels: transition rates from the first excited states to the ground state also as a function of $d/\gamma_0$. In all these plots we have fixed $d_0 =0.5\gamma_0 $ and  $N=2$, panels (a) and (b), and $N=4$, panels (c) and (d), TLS's coupled to the main one.}
\label{fig:energies_transitionrates}
\end{figure} 
When the coupling between the main and the first TLS is of the order or weaker than its decay rate, i.e., when $d_0 \lesssim \gamma_0$, the system can present interference between the different excitation paths, this regime is represented by the gray area showed \ref{fig:energies_transitionrates}(a) and (c). Otherwise, for $d_0 \gg \gamma_0$ the separation of the energy levels can be large enough to produce Autler-Townes splitting. On the left panels of Fig. \ref{fig:energies_transitionrates} we plot the first eigenenergies (ground and eigenstates with one excitation) of the system as a function of $ d/\gamma_0$, keeping $\Omega_p = 0$ and $d_0/\gamma_0 = 0.5\gamma_0$ (see expressions for the eigenstates/eigenenergies in the Supplementary Material for the case $N=2$). From this figure we can see, for the first region of parameters ($d < \gamma_0$) all the energy levels are within the linewidth of the excited state of the first TLS ($\gamma_0$); otherwise in the intermediate region ($d \sim \gamma_0$) some levels can be inside and others outside the linewidth of the excited state, thus presenting close eigenstates with possibly different decay rates. Finally, for very strong couplings ($d_0$ and $d$ much stronger than $\gamma_0$) is observed a complete level splitting. Thus, depending on the set of parameters, the system display total interference between the excitation paths (EIT \cite{Fleischhauer2005}), close different resonant states with asymmetric line-shape (Fano interference \cite{Fano1961}), or a complete separation of the levels (Autler-Townes splitting \cite{Fleischhauer2005}).

The decay rates of the excited eigenstates $|\psi_{k}\rangle$ of the whole system to its ground state $|\psi_{g}\rangle$ can be calculated via the Fermi's golden rule \cite{Cohen-Tannoudji1992}
\begin{equation}
\Gamma_{kg} = \gamma_0 |\langle \psi_{g}| \sigma _{-}^{0}| \psi_{k}\rangle|^2,
\end{equation}
where we have neglected the dissipation channels related to the other TLS 's we are assuming $\gamma_0 \gg \gamma$. For a few TLS's we can analytically derive the eigenstates and then the analytical expressions for the decay rates (See Supplementary Material).
In the Figs.\ref{fig:energies_transitionrates}(b) ($N=2$) and \ref{fig:energies_transitionrates}(d) ($N=4$) are plotted the transitions rates between the excited and ground states of the whole system as a function of the coupling $d$, for a given coupling $d_0$. As can be seen these decay rates are always different, except for a specific value of $d$, where all the transition rates coincide ($d=d_0/\sqrt{2}$). Such feature will have a direct effect on the optical properties of the system as it will discussed bellow.   

\begin{figure}[t]
\centering
\includegraphics[width=0.5\textwidth]{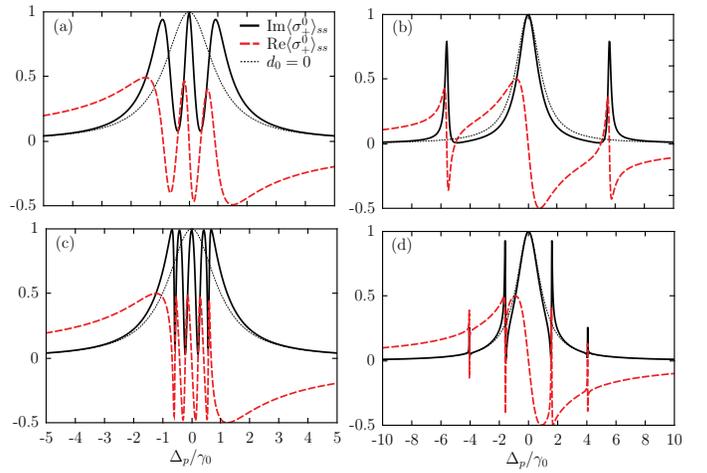}
\caption{Normalized absorption, $\text{Im}\langle \sigma_{+}^0 \rangle_{ss}$, (black solid line) and dispersion, $\text{Re}\langle \sigma_{+}^0 \rangle_{ss}$, (red dashed line) of the first TLS when coupled to $N$ TLS's as a function of the normalized detuning $\Delta_p /\gamma_0$: (a) and (b) for $N=2$, and (c) and (d) for $N=4$. The parameters used here were $\Omega_p = 0.03 \gamma_0$, $\gamma_i= \gamma = 10^{-3}\gamma_0$. In (a) we have chosen $d_0 =0.8 \gamma_0$ and $d =0.4 \gamma_0$, which allow us to observe two transparency windows. In (b), by choosing $d_0 = 2.5 \gamma_0$ and $d = 5 \gamma_0$ we have an Autler-Townes splitting. (c) Setting the parameters $d_0 = 0.5 \gamma_0$ and $d = d_0/\sqrt{2}$ we have multiple transparency windows, all of them with the same depth and width since we have adjusted the coupling $d$ with the value which allows the same transition rate from the excited states to the ground one of the whole system. Finally, in (d) we have fixed the parameters $d_0 = 0.5 \gamma_0$ and $d = 2.5 \gamma_0$, which allow for Fano interference. The black dotted lines represent the absorption when there is no TLS coupled to the main one.}
\label{fig:multi_FS}
\end{figure}

In Fig. \ref{fig:multi_FS} we plot the absorption and dispersion of the first TLS coupled to $N=2$, panels (a) and (b), or $N=4$, panels (c) and (d), other TLS's.  In the DDIT regime, the outer peaks are related to the coupling between the main and the second TLS (whose positions depend on $d_0$), while the inner peaks are related to the new resonant states introduced by the other coupled TLS's (whose position and width depend on $d$). In this way, the number of transparency windows is exactly equals to the number of TLS's ($N$) coupled to the main one.  For $d_0 < \gamma_0$, Figs. \ref{fig:multi_FS}(a) and (c), we have multiple transparency windows (multi-DDIT), while for $d_0$ and $d$ much stronger than $\gamma_0$ (Fig. \ref{fig:multi_FS}(b) we have a Autler-Townes splitting. For $d_0 < \gamma_0$ and $d>\gamma_0$ we have asymmetric excitation paths, resulting in resonant states with asymmetric line-shapes. This happens since by increasing the coupling $d$ the inner peaks become broader and then, depending on the coupling $d_0$, they can approach the other peaks, producing interference in the absorption paths, i.e., Fano interferences \cite{Fano1961}. The depth of the transparency windows is strongly dependent on the decay rate $\gamma$ (which works out as the dephasing rate in EIT experiments). In Fig. \ref{fig:multi_FS} all the depths of the transparency windows are close to the maximum value since we have assumed very small $\gamma$ (i.e, $\gamma = 0.001 \gamma_0$). On the other hand, the width of the transparency windows dependent on the transition rates between the excited and ground states of the whole system. As seen in Fig. \ref{fig:energies_transitionrates}(b) ($N=2$) and \ref{fig:energies_transitionrates}(d) ($N=4$) all the transitions rate are equal at a specific value $d=d_0/\sqrt{2}$. When they are equal, the width of the resonance peaks of the system, which also depends on the coupling $d$, are the same and then we end up with perfectly symmetric DDIT windows. 

Independent of the number $N$ of the TLS's coupled by the main one, the point where all the decay rates cross occurs always at $d=d_0/\sqrt{2}$. (We were able to derive the decay rates and the crossing points for up to $N=4$ TLS's coupled to the main one, as can be seen in the Supplementary Material). So, by choosing the specific parameters for that crossing point we will have a perfectly symmetrical absorption profile.
\begin{figure}[t]
\centering
\includegraphics[width=0.5\textwidth]{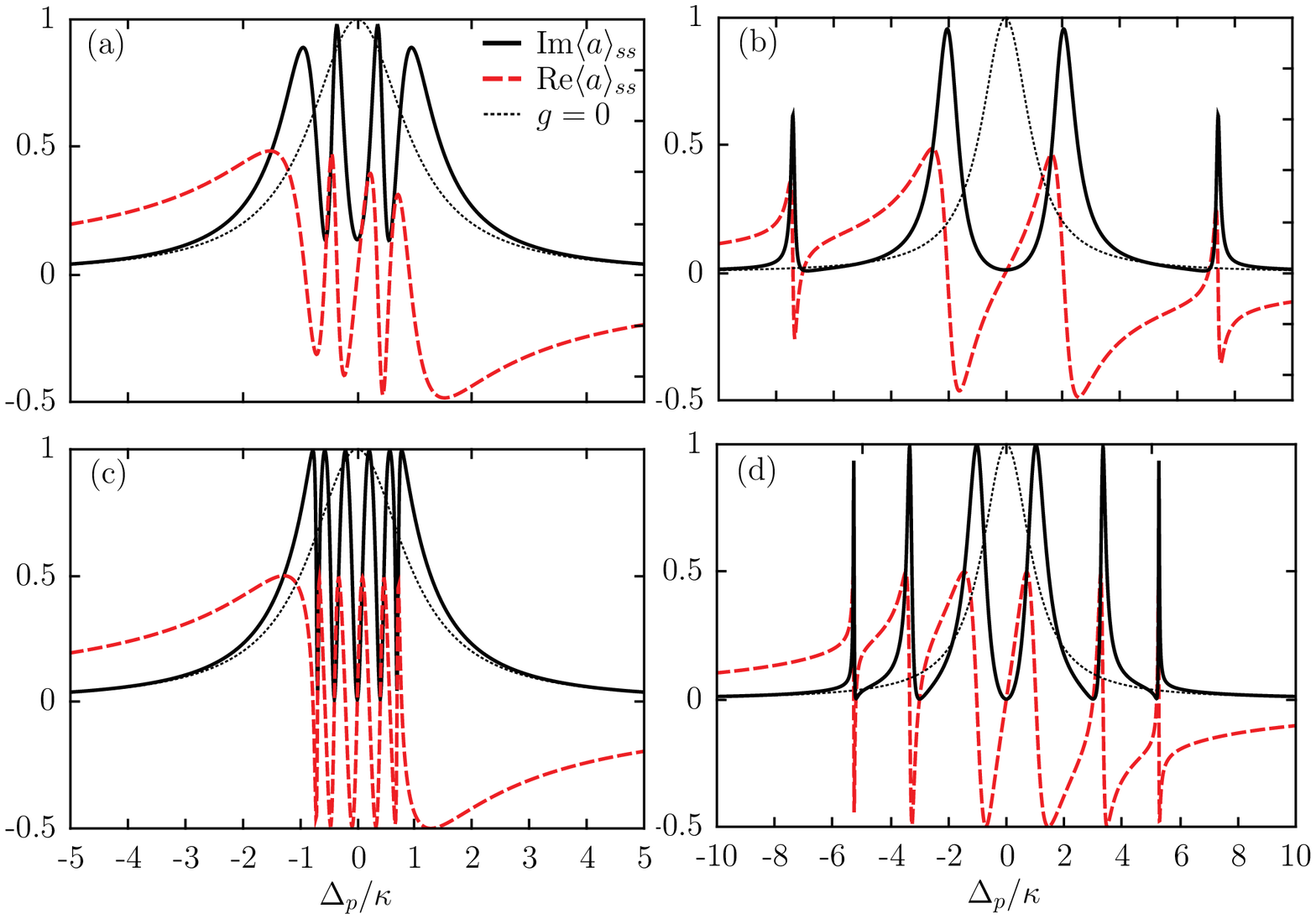}
\caption{Normalized absorption $\text{Im}\langle a \rangle_{ss}$ (black solid line) and dispersion $\text{Re} \langle a \rangle_{ss}$ (red dashed line) of the cavity mode when coupled to $3$, (a) and (b), and $5$, (c) and (d), two-level systems (i.e., the main TLS plus $N=2$ and $N=4$ other TLS's, respectively) as a function of the normalized detuning $\Delta_p /\kappa$. The parameters used here were $|\epsilon| = 0.03\kappa$, $\gamma_i= \gamma = 10^{-3}\kappa$. The other parameters were chosen to show: (a) two-transparency windows ($d = 0.4 \kappa$, $g = 0.8 \kappa$, and $\gamma_0 = 0.1 \kappa$); (b) Autler-Townes splitting and Fano interference ($d = 5.0 \kappa$, $g = 3.0 \kappa$, and $\gamma_0 = 0.1 \kappa$); (c) four identical transparency windows ($d = 0.4 \kappa$, $g = \sqrt{2}d$, and $\gamma_0 = 10^{-3} \kappa$); and finally (d) multiple Fano interferences ($d = 3.0 \kappa$, $g = 2.0 \kappa$, and $\gamma_0 = 10^{-3} \kappa$). The black dotted lines represent the absorption when there is no TLS coupled to the cavity mode.}
\label{fig:multi_Cav}
\end{figure} 

The multi-DDIT or multi-Fano interference also appear when we couple the series of $1+N$ TLS's to a cavity mode (as schematically shown in Fig.\ref{fig:Scheme}(b). As discussed above, the number of transparency windows is equal to the number of TLS coupled to the main one. Thus, considering $N$ TLS's coupled to the main one (each coupling given by $d$), which in turn is coupled to the cavity mode (coupling $g$), we will have $N$ transparency windows, as we see in Fig. \ref{fig:multi_Cav}(a) ($N=2$) and \ref{fig:multi_Cav}(c) ($N=4$), which present 2 and 4 inner peaks, respectively. The position of the resonance peaks is determined by all the couplings. However, the outer peaks are mainly due to the atom-field coupling $g$ and the inner peaks (and its widths) are mainly influenced by the dipole-dipole couplings $d$. For stronger values of $g$ and $d$ we can have a large separation between the resonance peaks (Autler-Townes splitting) or even Fano interference when $d>g$, as we see in Fig. \ref{fig:multi_Cav}(b) and \ref{fig:multi_Cav}(d). 

As we can see from Fig. \ref{fig:multi_Cav}, the number of transparency windows (inner peaks) is exactly equals to the number of TLS's coupled to the main one, as any TLS added to the system introduces a new resonant eigenstate (with one excitation). So, here we have a tunable system which allows us to arbitrarily choose the number of transparency windows, and their width, by simply adjusting the number of TLS's and the dipole-dipole coupling in our model .

\section*{Conclusions}
In summary, here we have investigated how dipole-dipole can induce transparency on a TLS or on a cavity mode. The dipole-dipole coupling works out as the control field in EIT or cavity-EIT experiments, while the decay rate of the first (second) TLS is the equivalent to the total decay rate of the excited state (dephasing rate of the ground state which is coupled to the excited one via control field) in EIT experiments with three-level atoms. Thus, we can identify a complete correspondence between the DDIT and the usual EIT in three-level atoms in $\Lambda $ configuration. We also could show the scalability of our system: by coupling more TLS's our system presents more transparency windows, being their number exactly equals to the number of TLS's coupled to the main one. The separation between the transparency windows and their widths depend on the dipole-dipole couplings and on the decay rate of the first excited states of the system turning it easily adjustable. We hope this new kind of induced transparency could be useful for manipulation of the optical properties of TLS in general, the study of slow light, transport properties in spin chains, and also frequency filter for light fields. Also, by detecting the optical response of a driven TLS/cavity mode we can estimate properties of dipole-dipole interaction.

\section*{Acknowledgments} 
C.J.V.-B. and H. S. Borges acknowledge support from CNPq and S\~{a}o Paulo Research Foundation (FAPESP), grant 2013/04162-5 and grant 2014/12740-1, and the Brazilian National Institute of Science and Technology for Quantum Information (INCT-IQ). C.J.V.-B. and E.C.D. acknowledge support from CNPq. 

\section*{Author contributions}
All the authors contributed substantially to this work.

\section*{Additional information}
Supplementary information is available in the online version of the paper. Correspondence and requests for materials should be
addressed to H. S. Borges.

\section*{Competing financial interests}
The authors declare no competing financial interests.

\section*{Methods}
\textbf{Obtaining the optical response}
In order to obtain analytically the optical response of a TLS coupled to linear a series of others $N$ TLS's, we derive a set of differential coupled equations    whereby in the steady regime ($\dot{\rho}=0$), provide us $\langle\sigma_+^0\rangle_{ss}$. To this end we assuming the Rabi frequency of the probe field much weaker than dipole coupling between the main TLS and its the first neighbour, i.e, $\Omega_p\ll d_0$ and its decay rate much stronger than the decay rate of the others one ($\gamma_0\gg\gamma_i$). With these assumptions we are able to use the semiclassical approximation, where the correlations between the atomic operator can be neglected, allowing us to solve the following set of equations:

{\footnotesize{
\begin{eqnarray*}
\left\langle \dot{\sigma}_{+}^{0}\right\rangle &=&\left(i\Delta _{p}-\gamma_0\right)\left\langle
\sigma _{+}^{0}\right\rangle -id_{0}\left\langle \sigma
_{z}^{0}\right\rangle \left\langle \sigma _{+}^{1}\right\rangle -i\Omega
_{p}\left\langle \sigma _{z}^{0}\right\rangle, \\
\left\langle \dot{\sigma}_{+}^{j}\right\rangle &=&\left(i\Delta _{p}-\gamma_j\right)\langle
\sigma _{+}^{j}\rangle -i\left\langle \sigma
_{z}^{j}\right\rangle \left[d_{j-1} \langle\sigma _{+}^{j-1}\rangle +d_{j+1}\langle \sigma _{+}^{j+1}\rangle\right]\\
& &\left(\text{for $1\leqslant j\leqslant N-1$}\right),\\
\left\langle \dot{\sigma}_{+}^{N}\right\rangle &=&\left(i\Delta_p-\gamma_N\right)
\left\langle \sigma _{+}^{N}\right\rangle -id_{N-1}\left\langle \sigma
_{z}^{N}\right\rangle \left\langle \sigma _{+}^{N-1}\right\rangle.
\end{eqnarray*}
}}\\
In the steady regime we end up with a system linear equations, whose solution was obtained using the function Solve of Mathematica software. Thus, for an arbitrary number of TLS's coupled to the main one we can derive analytically the expression of the $\langle \sigma_+^0\rangle$, allowing  us extract all the information respect to optical response of the system.

We also investigate here the optical properties of this scalable system when it has its main TLS coupled to a cavity mode. Analogously to the free space case, we obtain a set of linear coupled equations where we impose that the atomic and cavity field are not correlated. Thus, assuming the weak pumping field limit (whose strength is represented by $\epsilon$), i.e., $\epsilon\ll\kappa$ (being $\kappa$ the decay rate of the cavity field) we can derive the steady state solution for the  mean value of $\langle a\rangle_{ss}$ from the following system:
{\footnotesize{
\begin{eqnarray*}
\left\langle \dot{a}\right\rangle &=&-i(\Delta _{p}-i\kappa ) \left\langle a\right\rangle -ig\left\langle \sigma
_{-}^{0}\right\rangle -i\epsilon,\\
\left\langle \dot{\sigma}_{-}^{0}\right\rangle &=&-i\left(\Delta
_{p}-i\gamma_0\right)\left\langle \sigma _{-}^{0}\right\rangle +ig\left\langle
a\right\rangle \left\langle \sigma _{z}^{0}\right\rangle+id\left\langle
\sigma _{z}^{0}\sigma _{-}^{1}\right\rangle,\\
\left\langle \dot{\sigma}_{-}^{j}\right\rangle &=&-i\left(\Delta
_{p}-i\gamma_j\right)\left\langle \sigma _{-}^{j}\right\rangle +id\left\langle \sigma _{z}^{j}\sigma
\_^{j-1}\right\rangle +id\left\langle
\sigma _{z}^{j}\sigma \_^{j+1}\right\rangle \\
& &\left(\text{for $1\leqslant j\leqslant N-1$}\right),\\
\left\langle \dot{\sigma}_{-}^{N}\right\rangle &=&-i \left(\Delta_{p}-\gamma_N\right)\left\langle \sigma _{-}^{N}\right\rangle +id\left\langle \sigma _{z}^{N}\sigma \_^{N-1}\right\rangle.
\end{eqnarray*}
}}\\
We also used the Mathematica software to obtain the analytical expression of absorption and dispersion of the this system from the $\mathrm{Im}\langle a\rangle_{ss}$ and $\mathrm{Re}\langle a\rangle_{ss}$, respectively.    

\textbf{Data availability} The plots and results were developed using the analytical expressions that we derived. Any data that support the results showed within this manuscript are available from the corresponding author upon request.

\begin{widetext}

\textbf{Suplementary Information: Multi-Transparency Windows and Fano
interference Induced by Dipole-Dipole Couplings}

\begin{center}
E. C. Diniz, H. S. Borges, C. J. Villas-Boas
\end{center}

%% Text in onecolumn format.
\section{$1+N$ coupled TLS in Free Space}

In this first part we derive some expressions which appear or were used in
the main text regarding $1+N$ coupled TLS (with dipole-dipole interaction) in
free space.

\subsection{Derivation of the optical response for $N$ TLS coupled to the
main one} 
Here, we describe the derivation of the stationary solution of the optical
response of the main TLS, i.e., the average value $\langle \sigma_{+}^0
\rangle_{ss}=Tr(\rho_{ss}\sigma _{+}^{0})$.
We obtain the analytical solution in the steady state of an arbitrary number of TLS coupled to the main one. We assume weak probe field limit, i.e., $\left\vert
\Omega _{p}\right\vert $ $<<\left\vert d_{0}\right\vert $, and decay rate of
the main TLS stronger than that of the other, i.e., $(\gamma _{0}>>\gamma
_{1}).$

Such assumptions allow us to employ the the semiclassical approximation \cite{Carmichael}. From this approximation, we find the equations of motion for the expectation values of the system operators where the correlations between atomic operators are neglected, i.e, $\left \langle {\sigma_{\pm }^{i}} {\sigma_{z }^{j}}  \right \rangle\approx \langle {\sigma_{\pm }^{i}}\rangle\left \langle {\sigma_{z}^{j}} \right \rangle$. In low atomic excitation limit we assume, $\left \langle {\sigma_{z}^{i}}  \right \rangle\approx -1. $ Based on these assumptions we can derive the a general (arbitrary $N$) system of equations for the expectation values of the TLS operators 
\begin{eqnarray*}
\left\langle \dot{\sigma}_{+}^{0}\right\rangle &=&\left(i\Delta _{p}-\gamma_0\right)\left\langle
\sigma _{+}^{0}\right\rangle -id_{0}\left\langle \sigma
_{z}^{0}\right\rangle \left\langle \sigma _{+}^{1}\right\rangle -i\Omega
_{p}\left\langle \sigma _{z}^{0}\right\rangle, \\
\left\langle \dot{\sigma}_{+}^{j}\right\rangle &=&\left(i\Delta _{p}-\gamma_j\right)\langle
\sigma _{+}^{j}\rangle -id_{j-1}\left\langle \sigma
_{z}^{j}\right\rangle \langle \sigma _{+}^{j-1}\rangle -id_{j+1}\left\langle \sigma _{z}^{j}\right\rangle\langle \sigma _{+}^{j+1}\rangle \left(\text{for $1\leqslant j\leqslant N-1$}\right),\\
\left\langle \dot{\sigma}_{+}^{N}\right\rangle &=&\left(i\Delta_p-\gamma_N\right)
\left\langle \sigma _{+}^{N}\right\rangle -id_{N-1}\left\langle \sigma
_{z}^{N}\right\rangle \left\langle \sigma _{+}^{N-1}\right\rangle.
\end{eqnarray*}

In all cases described below, we found the stationary solution for the system of equations using the software \textit{Mathematica}~\cite{Mathe}.

\subsubsection{2 coupled TLS ($N=1$) in Free Space} 
We obtain the set of coupled equation motion for this case by setting $N=1$:
\begin{eqnarray*}
\left\langle \dot{\sigma}_{+}^{0}\right\rangle &=&\left(i\Delta _{p}-\gamma_0\right)\left\langle
\sigma _{+}^{0}\right\rangle -id_{0}\left\langle \sigma
_{z}^{0}\right\rangle \left\langle \sigma _{+}^{1}\right\rangle -i\Omega
_{p}\left\langle \sigma _{z}^{0}\right\rangle , \\
\left\langle \dot{\sigma}_{+}^{1}\right\rangle &=&\left(i\Delta _{p}-\gamma_1\right)\left\langle
\sigma _{+}^{1}\right\rangle -id_{0}\left\langle \sigma
_{z}^{1}\right\rangle \left\langle \sigma _{+}^{0}\right\rangle.
\end{eqnarray*}
The steady state analytical solution for this case is given by in Eq.(4) in the main document. 

\subsubsection{3 coupled TLS ($N=2$) in Free Space} 
The set of coupled motion equations for this case reads: 
\begin{eqnarray*}
\left\langle \dot{\sigma}_{+}^{0}\right\rangle &=&\left(i\Delta _{p}-\gamma_0\right)\left\langle
\sigma _{+}^{0}\right\rangle -id_{0}\left\langle \sigma
_{z}^{0}\right\rangle \left\langle \sigma _{+}^{1}\right\rangle -i\Omega
_{p}\left\langle \sigma _{z}^{0}\right\rangle , \\
\left\langle \dot{\sigma}_{+}^{1}\right\rangle &=&\left(i\Delta _{p}-\gamma_1\right)\left\langle
\sigma _{+}^{1}\right\rangle -id_{0}\left\langle \sigma
_{z}^{1}\right\rangle \left\langle \sigma _{+}^{0}\right\rangle
-id\left\langle \sigma _{z}^{1}\right\rangle \left\langle \sigma
_{+}^{2}\right\rangle ,
\\
\left\langle \dot{\sigma}_{+}^{2}\right\rangle &=&\left(i\Delta _{p}-\gamma_2\right)\left\langle
\sigma _{+}^{2}\right\rangle -id\left\langle \sigma _{z}^{2}\right\rangle
\left\langle \sigma _{+}^{1}\right\rangle.
\end{eqnarray*}
The stationary solution is: 
\begin{equation*}
\langle \sigma_+^0\rangle_{ss}=-\frac{\Omega_p \left[d^2-(\Delta_p+i \gamma_1) (\Delta_p+i \gamma_2)\right]}{(\Delta_p+i\gamma_0) \left[d^2-(\Delta_p+i \gamma_1) (\Delta_p+i \gamma_2)\right]+d_0^2 (\Delta+i \gamma_2)}
\end{equation*}%

\subsubsection{4 coupled TLS ($N=3$) in Free Space} 
For $N=3$ we similarly obtain the following system of equations:
\begin{eqnarray*}
\left\langle \dot{\sigma}_{+}^{0}\right\rangle &=&\left(i\Delta _{p}-\gamma_0\right)\left\langle
\sigma _{+}^{0}\right\rangle -id_{0}\left\langle \sigma
_{z}^{0}\right\rangle \left\langle \sigma _{+}^{1}\right\rangle -i\Omega
_{p}\left\langle \sigma _{z}^{0}\right\rangle , \\
\left\langle \dot{\sigma}_{+}^{1}\right\rangle &=&\left(i\Delta _{p}-\gamma_1\right)\left\langle
\sigma _{+}^{1}\right\rangle -id_{0}\left\langle \sigma
_{z}^{1}\right\rangle \left\langle \sigma _{+}^{0}\right\rangle
-id\left\langle \sigma _{z}^{1}\right\rangle \left\langle \sigma
_{+}^{2}\right\rangle ,
\\
\left\langle \dot{\sigma}_{+}^{2}\right\rangle &=&\left(i\Delta _{p}-\gamma_1\right)\left\langle
\sigma _{+}^{2}\right\rangle -id\left\langle \sigma
_{z}^{2}\right\rangle \left\langle \sigma _{+}^{1}\right\rangle
-id\left\langle \sigma _{z}^{2}\right\rangle \left\langle \sigma
_{+}^{3}\right\rangle ,\\
\left\langle \dot{\sigma}_{+}^{3}\right\rangle &=&\left(i\Delta _{p}-\gamma_3\right)\left\langle
\sigma _{+}^{3}\right\rangle -id\left\langle \sigma _{z}^{3}\right\rangle
\left\langle \sigma _{+}^{2}\right\rangle.
\end{eqnarray*}
The stationary solution is: 
\begin{equation*}
\langle \sigma^0_+\rangle_{ss}=\frac{i\Omega_p \left(i d^2 \left(\Delta_p+i \gamma_3\right)+\left(\gamma_1-i\Delta_p\right) \left(-d^2+\left(\Delta_p+i \gamma_2\right)
   \left(\Delta_p+i \gamma_3\right)\right)\right)}{d_0^2\left(-d^2+\left(\Delta_p+i\gamma_2\right)\left(\Delta_p+i \gamma_3\right)\right)+\left(\Delta_p+i \gamma_0\right) \left(d^2 \left(\Delta_p+i \gamma_3\right)-\left(\Delta_p+i \gamma_1\right) \left(-d^2+\left(\Delta_p+i\gamma_2\right) \left(\Delta_p+i \gamma_3\right)\right)\right)}
\end{equation*}%

\subsubsection{5 coupled TLS ($N=4$) in Free Space} 
Analogously, for $N=4$ we obtain the set of equations below: 
\begin{eqnarray*}
\left\langle \dot{\sigma}_{+}^{0}\right\rangle  &=&\left(i\Delta _{p}-\gamma_0\right)\left\langle
\sigma _{+}^{0}\right\rangle -id_{0}\left\langle \sigma
_{z}^{0}\right\rangle \left\langle \sigma _{+}^{1}\right\rangle -i\Omega
_{p}\left\langle \sigma _{z}^{0}\right\rangle, \\
\left\langle \dot{\sigma}_{+}^{1}\right\rangle &=&\left(i\Delta _{p}-\gamma_1\right)\left\langle
\sigma _{+}^{1}\right\rangle -id_{0}\left\langle \sigma
_{z}^{1}\right\rangle \left\langle \sigma _{+}^{0}\right\rangle
-id\left\langle \sigma _{z}^{1}\right\rangle \left\langle \sigma
_{+}^{2}\right\rangle,
\\
\left\langle \dot{\sigma}_{+}^{2}\right\rangle  &=&\left(i\Delta _{p}-\gamma_2\right)\left\langle
\sigma _{+}^{2}\right\rangle -id\left\langle \sigma _{z}^{2}\right\rangle
\left\langle \sigma _{+}^{1}\right\rangle -id\left\langle \sigma
_{z}^{2}\right\rangle \left\langle \sigma _{+}^{3}\right\rangle, \\
\left\langle \dot{\sigma}_{+}^{3}\right\rangle  &=&\left(i\Delta _{p}-\gamma_3\right)\left\langle
\sigma _{+}^{3}\right\rangle -id\left\langle \sigma _{z}^{3}\right\rangle
\left\langle \sigma _{+}^{2}\right\rangle -id\left\langle \sigma
_{z}^{3}\right\rangle \left\langle \sigma _{+}^{4}\right\rangle, \\
\left\langle \dot{\sigma}_{+}^{4}\right\rangle  &=&\left(i\Delta _{p}-\gamma_4\right)\left\langle
\sigma _{+}^{4}\right\rangle -id\left\langle \sigma _{z}^{4}\right\rangle
\left\langle \sigma _{+}^{3}\right\rangle.
\end{eqnarray*}

The solution in the steady state for this case is given by:
{\small 
\begin{equation*}
\langle \sigma_+^0\rangle_{ss}=\frac{i \Omega_p \left[d^2 \left(-d^2+\left(\Delta_p+i \gamma _3\right)
   \left(\Delta_p+i \gamma _4\right)\right)+\left(\Delta_p+i \gamma _1\right)
   \left(d^2 \left(\Delta_p+i \gamma _4\right)+\left(-\Delta_p-i \gamma _2\right)
   \left(-d^2+\left(\Delta_p+i \gamma _3\right) \left(\Delta_p+i \gamma
   _4\right)\right)\right)\right]}{d_0^2 \left[i d^2 \left(\Delta_p+i \gamma
   _4\right)+\left(\gamma _2-i \Delta_p\right) \left(-d^2+\left(\Delta_p+i \gamma
   _3\right) \left(\Delta_p+i \gamma _4\right)\right)\right]+\Upsilon}
\end{equation*} 
}
where

{\small 
$\Upsilon =\left(\Delta_p+i\gamma _0\right) \left[d^2 \left(\left(\gamma _3-i \Delta_p\right) \left(\Delta_p+i \gamma _4\right)+i d^2\right)+\left(\Delta_p+i \gamma _1\right) \left(d^2\left(\gamma _4-i \Delta_p\right)+i \left(\Delta_p+i \gamma _2\right)\left(-d^2+\left(\Delta_p+i \gamma _3\right)\left(\Delta_p+i \gamma _4\right)\right)\right)\right]$.
}

We were able to derive the steady state analytical solutions for other number of coupled TLS (up to $N=20$, depending on the computational capabilities). However, as the expression are very large, we do not present them here. Just to illustrate, in Fig. \ref{fig:multi_FSs} we present the absorption spectrum for $N=7,10, 12$ and $15$. Note that the number of transparency windows is exactly equals to $N$. Also, note that the depth of the transparency windows are different in this figure. This is due to the the non-null decay rate $\gamma$ used here. By increasing the number of TLS coupled to the main one, we increase the number of transparency windows. However, the higher the number of transparency windows, the more sensitive the system is to the noisy effects.
\begin{figure}[h]
\centering
\includegraphics[width=0.9\textwidth]{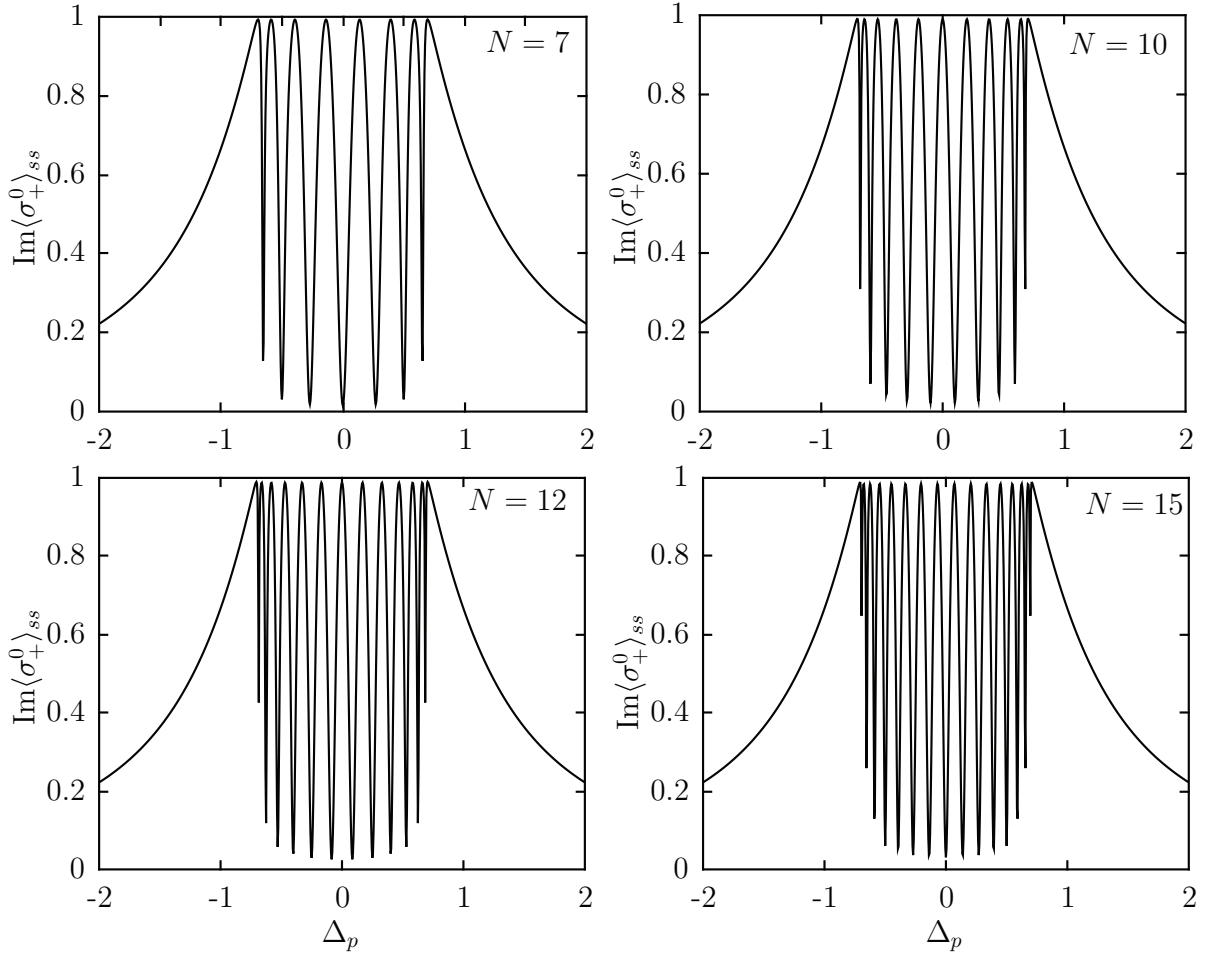}
\caption{Normalized absorption, $\text{Im}\langle \sigma_{+}^0 \rangle_{ss}$ as a function of the normalized detuning $\Delta_p /\gamma_0$, for different numbers of TLS coupled to the main one ($N=7,10, 12$ and $15$). The parameters used here were $\gamma_0=1$, $\Omega_p = 0.03 \gamma_0$, $\gamma_i= \gamma = 10^{-3}\gamma_0$, $d_0 = 0.5 \gamma_0$ and $d = d_0/\sqrt{2}$.}
\label{fig:multi_FSs}
\end{figure}

\subsection{Eigenstates, eigenenergies and transition rates between the
first excited states to the ground state for $1+N$ TLS}

The main goal of the present section is to present the analytical solutions
for the transition rates between the eigenstates of the system (from the
first excited one to the ground state). To this end, firstly we must derive
the eigenstates of our system, once the transition rate is defined as

\begin{equation}
\Gamma_{kg}= \gamma_0|\langle \psi_{g}| \sigma _{-}^{0}|
\psi_{k}\rangle|^2,
\end{equation}
being $|\psi_g\rangle$ and $|\psi_k\rangle$ ($k=1,2,...$) the ground and excited
eigenstates, respectively.

We have obtained analytically the expressions for the eigenstates and
transition rates for $N=2,3$ and $4$ only. As the expressions for the
eigenstates/eigenenergies are too extent, below we present them only for the $%
N=2$ case.

For $N=2$, the eigenvalues and its respective eigenvectors are

\begin{equation*}
E_0=-3\Delta \rightarrow |\psi_0\rangle= |ggg\rangle
\end{equation*}
\begin{equation*}
E_1=-\Delta-\sqrt{d_0^2+d^2} \rightarrow |\psi_1\rangle= \frac{d_0}{\sqrt{%
2(d_0^2+d^2)}}|egg\rangle -\frac{1}{\sqrt{2}}|geg\rangle + \frac{d}{\sqrt{%
2(d_0^2+d^2)}}|gge\rangle
\end{equation*}
\begin{equation*}
E_2=-\Delta \rightarrow |\psi_2\rangle= -\frac{d}{\sqrt{d_0^2+d^2}}%
|egg\rangle + \frac{d_0}{\sqrt{d_0^2+d^2}}|gge\rangle
\end{equation*}
\begin{equation*}
E_3=-\Delta+\sqrt{d_0^2+d^2} \rightarrow |\psi_3\rangle= \frac{d_0}{\sqrt{%
2(d_0^2+d^2)}}|egg\rangle +\frac{1}{\sqrt{2}}|geg\rangle + \frac{d}{\sqrt{%
2(d_0^2+d^2)}}|gge\rangle
\end{equation*}
\begin{equation*}
E_4=\Delta-\sqrt{d_0^2+d^2} \rightarrow |\psi_4\rangle= \frac{d}{\sqrt{%
2(d_0^2+d^2)}}|eeg\rangle -\frac{1}{\sqrt{2}} |ege\rangle + \frac{d_0}{\sqrt{%
2(d_0^2+d^2)}} |gee\rangle
\end{equation*}
\begin{equation*}
E_5=\Delta \rightarrow |\psi_5\rangle= -\frac{d_0}{\sqrt{d_0^2+d^2}}%
|eeg\rangle + \frac{d}{\sqrt{d_0^2+d^2}}|gee\rangle
\end{equation*}
\begin{equation*}
E_6=\Delta+\sqrt{d_0^2+d^2} \rightarrow |\psi_6\rangle= \frac{d}{\sqrt{%
2(d_0^2+d^2)}}|eeg\rangle +\frac{1}{\sqrt{2}} |ege\rangle + \frac{d_0}{\sqrt{%
2(d_0^2+d^2)}} |gee\rangle
\end{equation*}
\begin{equation*}
E_7=3\Delta \rightarrow |\psi_7\rangle= |eee\rangle.
\end{equation*}

With those eigenstates we can derive the transition rates between the first
excited states (with one excitation) and the ground state, which reads

\begin{eqnarray*}
\Gamma_{2g}=\frac{d^2}{d_0^2+d^2};\qquad \Gamma_{1g}=\Gamma_{3g}=\frac{d_0^2}{%
2\left(d^2+d_0^2\right)},
\end{eqnarray*}

From these expressions we find that these transition rates have a crossing
point at $d=\frac{d_0}{\sqrt{2}}$.

For $N=3$, the transition rates are given by:

\begin{eqnarray*}
\Gamma_{2g}=\Gamma_{3g}=\frac{2d^2-d_0^2+\sqrt{4d^2+d_0^2}}{4\sqrt{4d^2+d_0^2}}%
;\qquad \Gamma_{1g}=\Gamma_{4g}=\frac{-2d^2+d_0^2+\sqrt{4d^2+d_0^2}}{4\sqrt{%
4d^2+d_0^2}},
\end{eqnarray*}
and the crossing point of the transition rates is exactly the same, $d=\frac{%
d_0}{\sqrt{2}}$.

For $N=4$, the expressions of transitions rates follow below, 
{\normalsize 
\begin{eqnarray}
\begin{split}
\Gamma_{3g}=&\frac{d^2}{d^2+2d_0^2}, \\
\Gamma_{2g}=&\Gamma_{4g}=\frac{d_0^2\left(2d^2-d_0^2+C\right)}{2\left(d^2+2d_0^2%
\right)C},  \notag \\
\Gamma_{1g}=&\Gamma_{5g}=\frac{d_0^2\left(-2d^2+d_0^2+C\right)}{2\left(d^2+2d_0^2%
\right)C},
\end{split}%
\end{eqnarray}
} where $C=\sqrt{5d^4-2 d^2 d_0^2+d_0^4}$. For this configuration we have found two crossing points: $d=\frac{d_0}{\sqrt{2}}$, in which all the rates cross and $d=\sqrt{\frac{2}{5}}d_0$, where some rates cross.

\section{DDIT IN CIRCUIT QED: Analytical solutions}

Here we consider $1+N$ coupled TLS with the first one interacting with a
cavity mode, for instance in the Circuit QED framework.

\subsection{Derivation of the optical response for $1+N$ TLS coupled to a
cavity mode}

Employing the semiclassical approximation \cite{Carmichael} that allows to
factorize the correlator $\left\langle a\sigma_{-}^i\right\rangle \approx
\langle \sigma _{-}^i\rangle \left\langle a\right\rangle $, we
obtain the analytical solution for the average value of the annihilation
operator of the cavity mode for $1+N$ coupled TLS case. This semiclassical
approach is a good approximation whenever the driving field is very weak
compared to the dissipation rate of the cavity mode and the atom-field
coupling is also not so strong (again, when compared to the cavity field
decay rate $\kappa$).

The master equation for $1+N$ TLS coupled to a cavity mode, described in the
main document, is given by 
\begin{equation*}
\dot{\rho}=-i\left[ H_{c},\rho \right] +\overset{N}{\underset{i=0}{\sum }}%
\gamma _{i}(2\sigma _{-}^{i}\rho \sigma _{+}^{i}-\sigma _{+}^{i}\sigma
_{-}^{i}\rho -\rho \sigma _{+}^{i}\sigma _{-}^{i})+\kappa (2a\rho a^{\dagger
}-a^{\dagger }a\rho -\rho a^{\dagger }a),
\end{equation*}%
with $H_c$ given by Eq. (5) of the main document.

In order to get obtain analytically the expected value of the annihilation operator in the steady state $\left\langle a\right\rangle_{ss}$, we assume $\left\langle\sigma _{z}^{i}\right\rangle\approx-1$, which is a good approximation whenever the average number of photons inside the cavity is sufficiently small, which can be achieved by driving the cavity mode with a weak probe field. 

The derivation of the steady state solution for $\langle a \rangle$ for arbitrary $1+N$ TLS's coupled to a cavity mode follows the recurrence relations for the equations of motion for the average value of the atomic/cavity mode operators: 
\begin{eqnarray*}
\left\langle \dot{a}\right\rangle &=&-i(\Delta _{p}-i\kappa ) \left\langle a\right\rangle -ig\left\langle \sigma
_{-}^{0}\right\rangle -i\epsilon,\\
\left\langle \dot{\sigma}_{-}^{0}\right\rangle &=&-i\left(\Delta
_{p}-i\gamma_0\right)\left\langle \sigma _{-}^{0}\right\rangle +ig\left\langle
a\right\rangle \left\langle \sigma _{z}^{0}\right\rangle+id\left\langle
\sigma _{z}^{0}\sigma _{-}^{1}\right\rangle,\\
\left\langle \dot{\sigma}_{-}^{j}\right\rangle &=&-i\left(\Delta
_{p}-i\gamma_j\right)\left\langle \sigma _{-}^{j}\right\rangle +id\left\langle \sigma _{z}^{j}\sigma
\_^{j-1}\right\rangle +id\left\langle
\sigma _{z}^{j}\sigma \_^{j+1}\right\rangle, (\text{for $1\leqslant j\leqslant N-1$})\\
\left\langle \dot{\sigma}_{-}^{N}\right\rangle &=&-i \left(\Delta_{p}-\gamma_N\right)\left\langle \sigma _{-}^{N}\right\rangle +id\left\langle \sigma _{z}^{N}\sigma \_^{N-1}\right\rangle.
\end{eqnarray*}

From now we will provide in the next subsections the stationary solution of $\left \langle a \right\rangle_{ss}$ considering different numbers of TLS coupled to the cavity mode.  

\subsubsection{2 TLS ($N=1$) coupled to a cavity mode}

Writing the time derivatives for the expectation values of $\left\langle
a\right\rangle $ and $\left\langle \sigma _{\pm }^{i}\right\rangle $ we
obtain the following equations for 2 TLS (i.e, $N=1$) case: 
\begin{eqnarray*}
\left\langle \dot{a}\right\rangle &=&-i(\Delta _{p}-i\kappa )\left\langle
a\right\rangle -ig\left\langle \sigma _{-}^{0}\right\rangle \ -i\epsilon, \\
\left\langle \dot{\sigma}_{-}^{0}\right\rangle &=&-i(\Delta _{p}-i\gamma
_{0})\left\langle \sigma _{-}^{0}\right\rangle + ig\left\langle
a\right\rangle \left\langle
\sigma_z^0\right\rangle + id\left\langle  \sigma _{z}^{0}\sigma _{-}^{1}\right\rangle, \\
\left\langle \dot{\sigma}_{-}^{1}\right\rangle &=&-i\left( \Delta
_{p}-i\gamma _{1}\right) \left\langle \sigma _{-}^{1}\right\rangle
+ id\left\langle \sigma _{z}^{1}\sigma _{-}^{0}\right\rangle.
\end{eqnarray*}%

The expected value of the $\left\langle a\right\rangle $ in the steady state
($\dot{\rho}=0)$ calculate through the above equations is: 
\begin{equation*}
\left\langle a\right\rangle _{ss}=-\left\{ \frac{\epsilon (-d^{2}+\left(
\Delta _{p}-i\gamma _{0}\right) \left( \Delta _{p}-i\gamma _{1}\right) )}{%
-g^{2}\left( \Delta _{p}-i\gamma _{1}\right) -d^{2}(\Delta _{p}-i\kappa
)+\left( \Delta _{p}-i\gamma _{0}\right) \left( \Delta _{p}-i\gamma
_{1}\right) (\Delta _{p}-i\kappa )}\right\} .
\end{equation*}%
As mentioned in the main document, this result is very similar to the one
obtained in \cite{Souza2015s}.

\subsubsection{3 TLS ($N=2$) coupled to a cavity mode}

Now we present the derivation of the steady state solution for the average
value of the annihilation operator ($a$) when considering 3 TLS coupled to
the cavity mode. Similarly to the above procedure, we obtain the time
derivative of the expected values of the system operators: 
\begin{eqnarray*}
\left\langle \dot{a}\right\rangle &=&-i(\Delta _{p}-i\kappa ) \left\langle a\right\rangle -ig\left\langle \sigma
_{-}^{0}\right\rangle -i\epsilon, \\
\left\langle \dot{\sigma}_{-}^{0}\right\rangle &=&-i\left(\Delta
_{p}-i\gamma_0\right)\left\langle \sigma _{-}^{0}\right\rangle +ig\left\langle a\right\rangle
\left\langle \sigma _{z}^{0}\right\rangle +id\left\langle \sigma
_{z}^{0}\sigma _{-}^{1}\right\rangle, \\
\left\langle \dot{\sigma}_{-}^{1}\right\rangle &=&-i\left(\Delta
_{p}-i\gamma_1\right)\left\langle \sigma _{-}^{1}\right\rangle +id\left\langle \sigma
_{z}^{2}\sigma _{-}^{0}\right\rangle +id\left\langle \sigma _{z}^{1}\sigma
_{-}^{2}\right\rangle, \\
\left\langle \dot{\sigma}_{-}^{2}\right\rangle &=&-i\left(\Delta
_{p}-i\gamma_2\right)\left\langle \sigma _{-}^{2}\right\rangle +d\left\langle \sigma
_{z}^{2}\sigma _{-}^{1}\right\rangle.
\end{eqnarray*}%
In this case, the expected value to $\left\langle a\right\rangle $ in steady
state is given by:
{\footnotesize{
\begin{equation*}
\left\langle a\right\rangle _{ss}=-\left\{ -\frac{\epsilon \left(
d^{2}\left( -i\gamma _{2}+\Delta _{p}\right) +\left( -i\gamma _{0}+\Delta
_{p}\right) \left( d^{2}-\left( -i\gamma _{1}+\Delta _{p}\right) \left(
-i\gamma _{2}+\Delta _{p}\right) \right) \right) }{-g^{2}\left(
d^{2}-(-i\gamma _{1}+\Delta _{p})(-i\gamma _{2}+\Delta _{p})\right)
+(d^{2}(-i\gamma _{2}+\Delta _{p})+\left( -i\gamma _{0}+\Delta _{p}\right)
\left( d^{2}-\left( -i\gamma _{1}+\Delta _{p}\right) \left( -i\gamma
_{2}+\Delta _{p}\right) \right) )\left( -i\kappa +\Delta _{p}\right)}%
\right\}. 
\end{equation*}
}}

\subsubsection{4 TLS ($N=3$) coupled to a cavity mode}

Following what was done before, we obtain the set equations below:%
\begin{eqnarray*}
\left\langle \dot{a}\right\rangle  &=&-i(\Delta _{p}-i\kappa)\left\langle a\right\rangle -ig\left\langle \sigma
_{-}^{0}\right\rangle -i\epsilon, \\
\left\langle \dot{\sigma}_{-}^{0}\right\rangle&=&-i\left(\Delta-i\gamma_0\right)
_{p}\left\langle \sigma _{-}^{0}\right\rangle +ig\left\langle a\right\rangle
\left\langle \sigma _{z}^{0}\right\rangle +id\left\langle \sigma
_{z}^{0}\sigma _{-}^{1}\right\rangle, \\
\left\langle \dot{\sigma}_{-}^{1}\right\rangle &=&-i\left(\Delta
_{p}-i\gamma_1\right)\left\langle \sigma _{-}^{1}\right\rangle +id\left\langle \sigma
_{z}^{1}\sigma _{-}^{0}\right\rangle +id\left\langle \sigma _{z}^{1}\sigma
_{-}^{2}\right\rangle, \\
\left\langle \dot{\sigma}_{-}^{2}\right\rangle  &=&-i\left(\Delta_p-i\gamma_2\right)\left\langle \sigma _{-}^{2}\right\rangle +id\left\langle \sigma
_{z}^{2}\sigma _{-}^{1}\right\rangle +id\left\langle \sigma _{z}^{2}\sigma
_{-}^{3}\right\rangle, \\
\left\langle \dot{\sigma}_{-}^{3}\right\rangle  &=&-i\left(\Delta
_{p}-i\gamma_3\right)\left\langle \sigma _{-}^{3}\right\rangle +id\left\langle \sigma
_{z}^{3}\sigma _{-}^{2}\right\rangle.
\end{eqnarray*}%
Using the above equations we obtain to steady state the solution to field
operator {\footnotesize 
\begin{equation*}
\left\langle a\right\rangle _{ss}=-\left\{ \frac{-\epsilon (-d^{2}\left(
d^{2}-(-i\gamma_{2}+\Delta _{p})(-i\gamma_{3}+\Delta _{p})\right) +\left(
-i\gamma _{0}+\Delta _{p}\right) (d^{2}(-i\gamma _{3}+\Delta _{p})+(-i\gamma
_{1}+\Delta _{p})\left( d^{2}-(-i\gamma _{2}+\Delta _{p})(-i\gamma
_{3}+\Delta _{p})\right) )))}{(-g^{2}(d^{2}(-i\gamma _{3}+\Delta
_{p})+(-i\gamma _{1}+\Delta _{p})(d^{2}-(-i\gamma _{2}+\Delta _{p})(-i\gamma
_{3}+\Delta _{p})))+\Psi _{a}}\right\} ,
\end{equation*}%
} with {\footnotesize 
\begin{equation*}
\Psi _{a}=(-d^{2}(d^{2}-(-i\gamma _{2}+\Delta _{p})(-i\gamma _{3}+\Delta
_{p}))+\left( -i\gamma _{0}+\Delta _{p}\right) (d^{2}(-i\gamma _{3}+\Delta
_{p})+(-i\gamma _{1}+\Delta _{p})(d^{2}-(-i\gamma _{2}+\Delta _{p})(-i\gamma
_{3}+\Delta _{p}))))(-i\kappa +\Delta _{p})).
\end{equation*}%
}

\subsubsection{5 TLS ($N=4$) coupled to a cavity mode} 

The motion equations for this case are
\begin{eqnarray*}
\left\langle \dot{a}\right\rangle&=&-i(\Delta _{p}-i\kappa )\left\langle a\right\rangle -ig\left\langle \sigma
_{-}^{0}\right\rangle -i\epsilon, \\
\left\langle \dot{\sigma}_{-}^{0}\right\rangle  &=&-i\left(\Delta_p-i\gamma_0\right)\left\langle
\sigma _{-}^{0}\right\rangle +ig\left\langle a\right\rangle \left\langle
\sigma _{z}^{0}\right\rangle +id\left\langle \sigma _{z}^{0}\sigma
_{-}^{1}\right\rangle\\
\left\langle \dot{\sigma}_{-}^{1}\right\rangle  &=&-i\left(\Delta
_{p}-i\gamma_1\right)\left\langle \sigma _{-}^{1}\right\rangle +id\left\langle \sigma
_{z}^{1}\sigma _{-}^{0}\right\rangle +id\left\langle \sigma _{z}^{1}\sigma
_{-}^{2}\right\rangle, \\
\left\langle \dot{\sigma}_{-}^{2}\right\rangle  &=&-i\left(\Delta
_{p}-i\gamma_2\right)\left\langle \sigma _{-}^{2}\right\rangle +id\left\langle \sigma
_{z}^{2}\sigma _{-}^{1}\right\rangle +id\left\langle \sigma _{z}^{2}\sigma
_{-}^{3}\right\rangle\\
\left\langle \dot{\sigma}_{-}^{3}\right\rangle  &=&-i\left(\Delta
_{p}-i\gamma_3\right)\left\langle \sigma _{-}^{3}\right\rangle +id\left\langle \sigma
_{z}^{3}\sigma _{-}^{2}\right\rangle +id\left\langle \sigma _{z}^{3}\sigma
_{-}^{4}\right\rangle\\
\left\langle \dot{\sigma}_{-}^{4}\right\rangle  &=&-i\left(\Delta
_{p}-i\gamma_4\right)\left\langle \sigma _{-}^{4}\right\rangle +id\left\langle \sigma
_{z}^{4}\sigma _{-}^{3}\right\rangle.
\end{eqnarray*}%

Again, by imposing $d\rho/dt = 0$, we can solve the above equations for the steady state regime. The solution is then given by
\begin{equation*}
\langle a \rangle _{ss}=-\frac{A}{B}
\end{equation*}%
with 
{\footnotesize 
\begin{equation*}
A =\left\{ d^{2}\epsilon (-i\gamma
_{4}+\Delta _{p})\left\{ -d^{2}+(-i\gamma _{0}+\Delta _{p})(-i\gamma
_{1}+\Delta _{p})+\left( -i\gamma _{0}+\Delta _{p}+\frac{(d^{2}-(-i\gamma
_{0}+\Delta _{p})(-i\gamma _{1}+\Delta _{p}))(-i\gamma _{2}+\Delta _{p})}{%
d^{2}}\right) (-i\gamma _{3}+\Delta _{p})-\Xi \right\} \right\} ,
\end{equation*}%
}
where
\begin{equation*}
\Xi =\frac{d^{2}\left( -i\gamma _{0}+\Delta _{p}+\frac{(d^{2}-(-i\gamma
_{0}+\Delta _{p})(-i\gamma _{1}+\Delta _{p}))(-i\gamma _{2}+\Delta _{p})}{%
d^{2}}\right) }{-i\gamma _{4}+\Delta _{p}},
\end{equation*}%
and
{\footnotesize
\begin{eqnarray*}
B &=&(d^{2}(-d^{2}(g^{2}-(-i\gamma
_{0}+\Delta _{p})\left( -i\kappa +\Delta _{p}\right) )+(-i\gamma _{2}+\Delta
_{p})(d^{2}\left( -i\kappa +\Delta _{p}\right) +(-i\gamma _{1}+\Delta
_{p})(g^{2}-(-i\gamma _{0}+\Delta _{p})\left( -i\kappa +\Delta _{p}\right)
)))- \\
&&(-i\gamma _{4}+\Delta _{p})(-d^{2}(d^{2}\left( -i\kappa +\Delta
_{p}\right) +(-i\gamma _{1}+\Delta _{p})(g^{2}-(-i\gamma _{0}+\Delta
_{p})\left( -i\kappa +\Delta _{p}\right) ) \\
&&+(-i\gamma _{4}+\Delta _{p})(-d^{2}(g^{2}-(-i\gamma _{0}+\Delta
_{p})\left( -i\kappa +\Delta _{c}\right) )+(-i\gamma _{2}+\Delta
_{p})(d^{2}\left( -i\kappa +\Delta _{p}\right) +(-i\gamma _{1}+\Delta
_{p})(g^{2}-(-i\gamma _{0}+\Delta _{p})\left( -i\kappa +\Delta _{p}\right)
))))).
\end{eqnarray*}%
}

Again, we were able to derive the steady state analytical solutions for other $N$'s (in this case, up to $15$), but the expression are very large to be presented here. Just to illustrate, in Fig. \ref{fig:multi_Cavs} we present the absorption spectrum for $N=7, 10, 12$ and $15$. Note that the number of transparency windows is exactly equals to $N$.
\begin{figure}[h]
\centering
\includegraphics[width=0.9\textwidth]{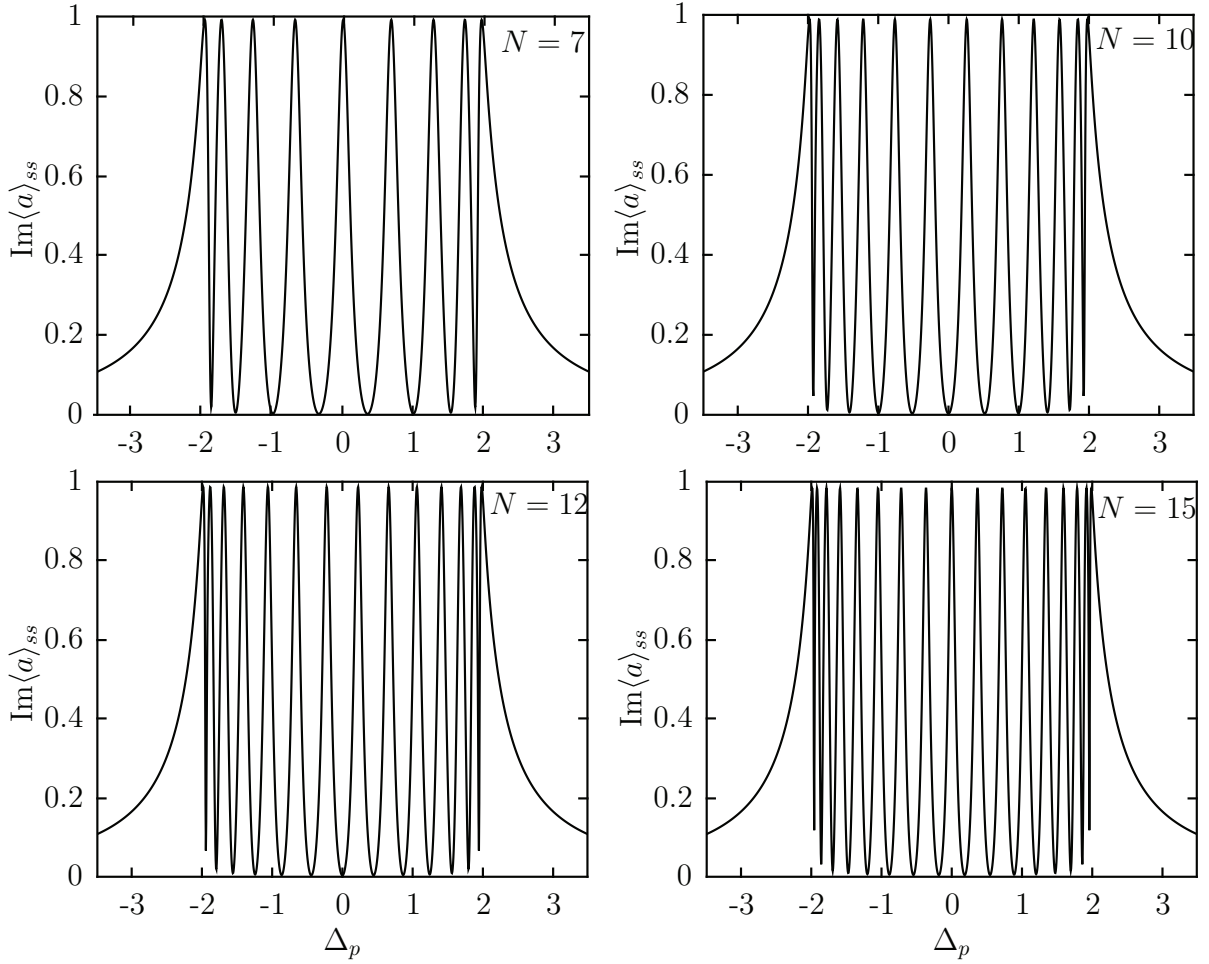}
\caption{Normalized transmission, $\text{Im}\langle a \rangle_{ss}$ as a function of the normalized detuning $\Delta_p /\gamma_0$, for different number of TLS $1+N$ coupled to the cavity mode ($N=7,10, 12$ and $15$). The parameters used here were $\kappa=1$, $\gamma_0=\gamma_i=10^{-3}\kappa$, $|\epsilon| = 0.03 \kappa$, $ d = 1.0\kappa$ and $g = \sqrt{2}d$.}
\label{fig:multi_Cavs}
\end{figure}

\subsection{Transition rates between the first excited states to the ground state for $1+N$ TLS coupled to a cavity
mode}

Analogously on free space, we could derive the transition rates for some cases when $1+N$ TLS are coupled to mode cavity. It can be calculated through the following expression:

\begin{equation*}
\Gamma_{kg}= \gamma_0|\langle \psi_{g}|a|
\psi_{k}\rangle|^2,
\end{equation*}
being $|\psi_g\rangle$ and $|\psi_k\rangle$ ($k=1,2,...$) the ground and excited
eigenstates, respectively.

For $N=2$, the rates are given by

\begin{eqnarray*}
\begin{split}
\Gamma_{1g}=&\Gamma_{4g}=\frac{d^2 g^2}{4d^4+g^4+\left(2d^2-g^2\right)\sqrt{%
4d^4+g^4}} \\
\Gamma_{2g}=&\Gamma_{3g}=\frac{d^2 g^2}{4d^4+g^4+\left(g^2-2d^2\right)\sqrt{%
4d^4+g^4}}. 
\end{split}%
\end{eqnarray*}
Similarly on it happens in the free space there is a crossing point associated to a specific value of $d$ where all the transition rates cross. For this case the crossing  point is $d =g/\sqrt{2}$.

For $N=3$, the rates are given by:

\begin{equation*}
\begin{split}
\Gamma _{3g}=& \frac{d^{2}}{d^{2}+2g^{2}} \\
\Gamma _{1g}=& \Gamma _{5g}=\frac{g^{2}\left( g^{2}-2d^{2}+C\right) }{2C\left(
d^{2}+2g^{2}\right) } \\
\Gamma _{2g}=& \Gamma _{4g}=\frac{g^{2}\left( 2d^{2}-g^{2}+C\right) }{2C\left(
d^{2}+2g^{2}\right) },
\end{split}%
\end{equation*}%
being $C=\sqrt{5d^{4}-2d^{2}g^{2}+g^{4}}$. For this case all the rates cross again at $d =g/\sqrt{2}$, while some rates cross at $d=\sqrt{2/5}g$.

\end{widetext}

\end{document}